\documentclass[aps,nofootinbib,preprint,showkeys]{revtex4}
\usepackage{graphicx}
\usepackage{amsfonts}

\newcommand{\fft}[2]{{\frac{#1}{#2}}}

\newcommand{\Tr}{{\rm Tr\,}}

\begin{document}

%\preprint{hep-th/yymmnnn}
\title{2D Heisenberg model from rotating membrane}

\author{Wen-Yu Wen}\email{steve.wen@gmail.com}
\affiliation{Department of Physics and Center for Theoretical
Sciences, National Taiwan University, Taipei 106, Taiwan}

%\date{\today}

\begin{abstract}
We study a rotating probe membrane in $S^3$ inside $AdS_4\times
S^7$ background of M-theory. With (partial) gauge fixing, we show
that in the {\it fast} limit the worldvolume of tensionless
membrane reduces to either the XXX$_{1/2}$ spin chain or the
two-dimensional $SU(2)$ Heisenberg spin model.  Later we introduce
the anisotropy and couple it to the external magnetic field. We
also establish the correspondence for higher dimensional
(D)$p$-branes.
\end{abstract}



\maketitle

\section{Introduction}

The anti-de Sitter/conformal field theory (AdS/CFT) correspondence
has revealed deep relation between string theory and gauge theory,
in particular the correspondence between IIB strings on
$AdS_5\times S^5$ and ${\cal N}=4$ super Yang-Mills theory (SYM)
at the 't Hooft's large N
limit\cite{Maldacena:1997re,Gubser:1998bc,Witten:1998qj}. While
the correspondence was mostly tested for those
Bogomol'nyi-Prasad-Sommerfield (BPS) states where partial
supersymmetry protects it against quantum
correction\cite{Aharony:1999ti}, semiclassical analysis on the
near-BPS sector is also considered in the
Ref.\cite{Berenstein:2002jq}; the integrability of this
Berenstein-Madalcena-Nastase (BMN) limit allows for quantitative
tests of correspondence beyond BPS states, where the energy of
classical string solutions is compared to the anomalous dimension
of SYM operators with large {\it R}-charge.  On the other hand,
intimate relation between SYM dynamics and integrable spin chain
was realized in the Ref.\cite{Beisert:2003ea}, because the planar
limit of the dilatation operator was identified with the
Hamiltonian of integrable spin chains.

The string/spin chain correspondence was first demonstrated in the
Ref.\cite{Kruczenski:2003gt}, where a classical rotating string on
$S^3$ inside $S^5$ was identified with the semiclassical coherent
state in the $SU(2)$ Heisenberg spin chain model. Later this
identification was explored in the full $SU(3)$
sector\cite{Hernandez:2004uw} and $SL(2)$
sector\cite{Bellucci:2004qr}.  A few examples of generalization
were discussed in the past: the fast spinning string in the
marginally deformed $AdS_3\times S^3$\cite{Israel:2004vv} and
$\beta$-deformed ${\cal N}=4$\cite{Lunin:2005jy}, each corresponds
to the anisotropic XXZ spin chain\cite{Wen:2006fw,Frolov:2005ty}.
The Melvin's magnetic-deformed background was also studied in the
Ref.\cite{Huang:2006bh}.

Strings, however, are not the only {\it elementary} objects since
we have learnt that there are more extended objects such as
D(Dirichlet)$p$-branes in the string theory as well as membranes
and five-branes in the M-theory.  Those extended objects are
soliton-like objects in the low energy description of
supergravity, resisting any naive perturbative analysis applicable
to strings.  Nevertheless, they may serve as a good probe to
explore non-perturbative aspects of string/M-theory, complementary
to our perturbative knowledge based on strings.  In addition to
their non-perturbative nature, there is no unique way to gauge
away degrees of freedom in the worldvolume of $p$-branes such as
what we usually do to the worldsheet of strings.  Despite of these
apparent difficulties, it is still an educated guess that there
might be similar correspondence between those extended objects and
higher dimensional spin systems in some specific limit.  In this
paper, we take the first step to support this conjecture with
concrete examples.  In particular, with careful choice of
lagrangian multipliers and proper gauge fixing, we are able to
achieve the correspondence between rotating membrane and
two-dimensional ($2$D) Heisenberg model and later generalize to
arbitrary $p$ dimensions. The outline of this paper is as follows:
In the section 2, a rotating tensionless membrane with gauge
fixing is shown to give rise to the $2$D Heisenberg model. The
condition of integrability for this spin model is briefly
mentioned. The anisotropy is introduced via the same deformation
as shown in the Ref.\cite{Wen:2006fw} and vortices-like excitation
is discussed. At the end, we show that upon a partial gauge fixing
the action can also reduce to the XXX$_{1/2}$ spin chain.  In the
section 3, the correspondence between $p$-branes and
higher-dimensional spin models is also established.  In
particular, for D$p$-branes there exist some new features due to
nontrivial dilaton field and two-form flux.  We also show that the
external magnetic field can be generated $\it geometrically$ in
all the $SU(2)$ spin models. In the conclusion, we summarize our
results and comment on possible directions for future
investigation. In the appendix we review the double scaling limit
constructed in the Ref.\cite{Wen:2006fw} for the anisotropic
Heisenberg model.

\section{2D Heisenberg model/membrane correspondence}
\subsection{$p$-brane action}
The Probe branes approach is widely used in the string/M-theory to
study different aspects of brane itself or specific background,
from classical to quantum levels.  It takes the assumption that
the back-reaction of the brane on the background is negligible. In
this paper we only study a single brane in the tensionless limit,
thus qualifying the probing assumption.  The brane action in
general falls into two types: the Dirac-Nambu-Goto type action and
the Polyakov type action.  The former one is intuitively simple
but carries nonlinearity due to the square root.  The latter one
introduces non-dynamical Lagrangian multipliers in order to
linearize the former action.  It is, however, easier for the
purpose of calculation and suitable for taking the tensionless
limit. To be specific, we adopt a Polyakov type action proposed
for general $p$-branes\cite{Bozhilov:2002sj}:
\begin{equation}\label{p-action}
S_p=\int{d^2\xi}\{
\fft{1}{4\lambda^0}[G_{00}-2\lambda^jG_{0j}+\lambda^i\lambda^jG_{ij}-(2\lambda^0T_p)^2|G_{ij}|]+T_p
b_{\mu_0\cdots\mu_p}\partial_0X^{\mu_0}\cdots\partial_{p}X^{\mu_p},
\}
\end{equation}
where $\vec{\xi}=\{\xi^0,...,\xi^p\}$ are worldvolume coordinates
and $\vec{X}=\{X^0,\cdots,X^D\}$ are ($D+1$)-dimensional target
spacetime. $\lambda$'s are the Lagrangian multipliers.
$i,j=\{1,...,p\}$ only run for spatial indices.
$G_{ij}\equiv\partial_i X^{\mu}\partial_jX^{\nu}g_{\mu\nu}$ is the
pull back metric on the worldvolume and $|G|$ is for determinant.
This action can be shown equivalent to the Dirac-Nambu-Goto type
action with the following equations of motion of $\lambda$'s
substituted in,
\begin{eqnarray}
&& G_{00}-2\lambda^jG_{0j}+\lambda^i\lambda^j G_{ij} +
(2\lambda^0T_p)^2 |G_{ij}|=0,\nonumber\\
&& G_{0j}-\lambda^iG_{ij}=0.
\end{eqnarray}

Without loss of generality, we may set $\lambda^i=0$ but
$\lambda^0\neq 0$.  Then the action is simplified and equations of
motion of $\lambda$'s reduce to the Virasora-like constraints
\begin{eqnarray}
&& G_{00} + (2\lambda^0 T_p)^2 |G_{ij}|=0,\\
&& G_{0j}=0.\label{eq:virasora}
\end{eqnarray}

\subsection{Gauge fixing and tensionless limit} We will restrict our discussion
on membranes in the M-theory for the moment and come back to
general p-branes in the next section. We first fix the gauge as
follows,
\begin{equation}
X^0=\kappa \xi^0,\qquad X^9=c_2\xi^2,\qquad X^{10}=c_1\xi^1,
\end{equation}
and the determinant decomposes into
\begin{equation}
|G_{ij}|=|\tilde{G}_{ij}|+c_2^2g_{99}\tilde{G}_{11}+c_1^2g_{[10][10]}\tilde{G}_{22}+c_1^2c_2^2g_{99}g_{[10][10]},
\end{equation}
where the $\tilde{G}_{ij}$ is the pull back metric of subspace
$\{X^1,\cdots,X^8\}$. Then the action becomes
\begin{eqnarray}
&&S_{m_2}=\int{d^3\xi}\{\fft{1}{4\lambda^0}[{G}_{00}-(2\lambda^0T_2)^2|\tilde{G}_{ij}|-{\rm
g}_1 ^2\tilde{G}_{11}-{\rm g}_2^2\tilde{G}_{22}-\fft{{{\rm
g}_1}^2{{\rm g}_2}^2}{2\lambda^0T_2}]+ T_2
b_{\mu\nu\lambda}\partial_0X^{\mu}\partial_1X^{\nu}\partial_2X^{\lambda}\},\nonumber\\
&&{\rm g}_1 \equiv 2\lambda^0 c_2 T_2\sqrt{g_{99}},\qquad {\rm
g}_2 \equiv 2\lambda^0 c_1 T_2\sqrt{g_{[10][10]}}.
\end{eqnarray}
In particular, we are looking for the limit where the membrane is
tensionless but effective couplings $\rm g_i$'s are finite.  This
can be achieved by sending $T_2\to0$ and $c_i\to \infty$ but
keeping their product finite.  With this gauge choice and scaling,
we only pay attention to the {\it local} property of the membrane
regardless its global topology. In the next section, we will see
that the tensionless membrane in this gauge fixing provides an
appropriate setting for the $2$D Heisenberg Model.

\subsection{2D Heisenberg model from fast membrane} In order to
reproduce the Heisenberg model, a fast membrane limit has to be
taken in analogy to the Ref.\cite{Kruczenski:2003gt}. We first
make the membrane rotate in one angular direction then take the
fast limit.  In this limit, we send $\kappa\to\infty$ and
$\dot{X}\to 0$ but keep $\kappa \dot{X}$ finite, here $\dot{X}$ is
partial derivative w.r.t. $\xi^0$.  The physics behind implies
that at this time scale, along those direction transverse to the
rotation plane, the membrane moves very slowly and can be seen as
almost frozen.  Although the conjugated energy and momentum at
this scale in fact blow up, the finite part of worldvolume action
is recognized as the sigma model of Heisenberg spin system. To
illustrate this, we will assume that a tensionless probe membrane,
constructed in the previous subsection, rotates inside a $S^3$,
which could be part of the eleven dimensional vacuum of M-theory,
such as $AdS_4\times S^7$.  The metric $g_{99},g_{[10][10]}$ are
assumed to take constant values on the sphere for simplicity
throughout the paper, but this assumption can be easily relaxed.
The relevant background metric after rotating, i.e. $\alpha\to
t+\alpha$, is given by
\begin{equation}\label{target_metric}
ds^2=\fft14[2dtd\alpha+d\alpha^2+d\beta^2+d\gamma^2+2\cos{\beta}d\alpha
d\gamma + 2\cos{\beta}dtd\gamma].
\end{equation}
After taking the fast limit, the finite worldvolume action with
pull back metric reads\footnote{The complete action also includes
a divergent constant term which is inversely proportional to
$T_2$. It is generically finite but seen as an artifact in the
tensionless limit.  Anyway, this constant term will not affect the
equations of motion.}
\begin{equation}
S_{m_2}=\fft{1}{16\lambda^0}\int{d^3\xi} \{
2\kappa\dot{\alpha}+2\kappa\cos{\beta}\dot{\gamma}-\sum_{i=1}^2{\rm
g}_i{}^2[(\partial_i\alpha)^2+(\partial_i\beta)^2+(\partial_i\gamma)^2+2\cos{\beta}\partial_i\alpha\partial_i\gamma]\}.
\end{equation}
Notice that only derivatives of first order in $\xi^0$ and second
order in $\xi^i$ are survived in this limit, which is the very
differential strucutre underlying the Heisenberg spin model.
Applying the constraint (\ref{eq:virasora}), one achieves the
action of Heisenberg model:
\begin{equation}\label{action_2D}
S_{m_2}=\fft{1}{16\lambda^0}\int{d^3\xi} \{ 2\partial_t\alpha+
2\cos{\beta}\partial_t\gamma -\sum_{i=1}^2{\rm
g}_i{}^2[(\partial_i\beta)^2+\sin^2{\beta}(\partial_i\gamma)^2]
\},
\end{equation}
or the Hamiltonian after the Legendre transformation,
\begin{equation}\label{hamil}
H_{2D}=\fft{\rm g^2}{16\lambda^0}\int{d\xi^1d\xi^2}
(\partial\vec{S})^2.
\end{equation}
Here we have made $\rm{g}_i=\rm{g}$ for the universal coupling and
defined a $SO(3)$ vector $\vec{S} =
(\sin{\beta}\sin{\gamma},\sin{\beta}\cos{\gamma},\cos{\beta})$,
which smoothly rotates as we move around the $\xi^1$-$\xi^2$
plane. Strictly speaking, this Hamiltonian is the long wavelength
or continuous limit of that of the discrete lattice model, where
the only interaction is among nearest neighbors. Therefore this is
straightforward generalization of one-dimensional spin chain to
higher dimensions. The nearest diagonal sites are next-nearest
neighbors and interaction among them only appears in the second
loop mixing.  In another words, that mixing enters the nonlinear
sigma model of worldsheet/worldvolume only via higher order
correction, if desired.

\subsection{Membrane excitation and integrability} In the
Ref.\cite{Bozhilov:2006bi}, magnon-like excitation was studied for
the membrane in the same Polyakov action. In our limit, the
membrane action is greatly simplified and excitation on the
membrane has its description in the corresponding Heisenberg
model.  The equation of motion derived from the
Eq.(\ref{action_2D}) is the very $(1+2)$-dimensional
Landau-Lifshitz equation, i.e.
\begin{equation}
\partial_t\vec{S}={\rm g}^2\vec{S}\times \partial^2\vec{S},
\end{equation}
which, different from its $(1+1)$-dimensional counterpart, is not
integrable in general.  This system is gauge equivalent to the
$(1+2)$-dimensional non-linear Schr\"{o}dinger-type
equation\cite{QingDing:2005}. It has been shown that, however, for
the following travelling wave ansatz it is still integrable,
\begin{equation}
\vec{S}(\vec{\xi},\xi^0)=\vec{S}(\vec{\xi}-\vec{v}\xi^0),\qquad
\vec{\xi}=(\xi^1,\xi^2).
\end{equation}
We will follow the Ref.\cite{Papanicolaou:1981cv} to discuss its
integrability by taking the advantage of isomorphism $SO(3)\simeq
SU(2)$.  To do so, we set $\rm g=1$ for simplicity and rewrite
$S=S^a\hat{\sigma}^a$, where $\hat{\sigma}^a$ are the Pauli
matrices and introduce complex variables $z=\fft12(\xi^1+i\xi^2)$
and $\eta=v^1+iv^2$, and their complex conjugates
$\bar{z},\bar{\eta}$.  Then one can form a Lax pair $(C^-,C^+)$
with a $2\times 2$ complex matrix $\Psi$,
\begin{eqnarray}
&&\partial_{z}\Psi=-C^-\Psi,\nonumber\\
&&\partial_{\bar{z}}\Psi=C^+\Psi,\nonumber\\
&&C^-\equiv
\frac{[S,\partial_zS]}{2i(x-i)}+\frac{x\bar{v}S}{(x-i)^2},\qquad
C^+\equiv
\frac{[S,\partial_{\bar{z}}S]}{2i(x+i)}+\frac{xvS}{(x+i)^2}.
\end{eqnarray}
The integrability condition thus reads
$C^+_z+C^-_{\bar{z}}-[C^+,C^-]=0$.  The arbitrary real spectral
parameter $x$ relates to a rotation on $v$ in the following way,
\begin{equation}
v'=e^{-i\theta}v,\qquad x=\pm\cot{\theta/2}.
\end{equation}
In fact, one can construct a new travelling wave solution with
velocity $\vec{v'}$
\begin{equation}
\vec{S'}=\vec{S'}(\vec{\xi}-\vec{v'}\xi^0),
\end{equation}
where $S'=\Psi^\dagger S \Psi$ with $S'=S'^{a}\hat{\sigma}^a$.

\subsection{Anisotropy and vortex dynamics} It has been shown in
the Ref.\cite{Wen:2006fw} that in the limit of fast string and
small deformation, a one-parameter deformation on the target $S^3$
reproduces the anisotropic Heisenberg spin chain $XXZ_{1/2}$. This
generalization is also applicable to the fast membrane considered
in the present paper.  We will summarize the derivation in the
Appendix and only present the result here.  The modified
Landau-Lifshitz equation is now with an anisotropic $3\times 3$
matrix ${\cal J}$, i.e.
\begin{eqnarray}\label{ll_aniso}
&&\partial_t \vec{S} = {\rm g}^2 \vec{S} \times \partial^2\vec{S}
+ \vec{S}\times
{\cal J}\vec{S},\nonumber\\
&&{\cal J}_{11}={\cal J}_{22}=1,\qquad {\cal J}_{33}=1-\delta.
\end{eqnarray}
$\delta$ is the deformed parameter which rescales the $U(1)$ fiber
in the Hopf fibration of $S^3$.  It reduces to $XY$-model and
isotropic $XXX_{1/2}$ model for $\delta=1$ and $\delta=0$,
respectively.  The same model has been applied to the study of
two-dimensional magnets for decades and we expect it could be
carried over, with some caution, to the study of tensionless
membrane in the fast/small deformation limit.  The excitation in
the two-dimensional system is much richer than that in the
one-dimensional spin chain. To our particular interests,
topological objects such as vortices can be
excited\cite{GWBM:1989,MB:1999}.  The vortices here carry two
topological charges under the homotopy group $\pi_1$ and $\pi_2$,
respectively.  They are the vorticity, $q=\pm1,\pm2,\cdots$ and
the polarization $p$, where $p=\pm1$ for non-planar excitation and
$p=0$ for planar one.  The dynamics of vortices is governed by
both the Coulomb force and {\it Magnus} force.  The latter is a
gyro force coupled to the vortex velocity, thus its behavior is
non-Newtonian.  A single static vortex is non-local for its naked
topological charge.  A vortex-antivortex pair, however, can be
localized in the form of dipole charge. Other than vortices, one
also expects the meson-like excitation, magnons, appears as usual.
A domain wall could even form if several of them are excited
coherently.

In fact, the shape of vortex can be given by a more generic
travelling wave ansatz\cite{MB:1999},
\begin{equation}
\vec{S}(\vec{\xi},\xi^0)=\vec{S}(\vec{\xi}-\vec{x},\dot{\vec{x}},\ddot{\vec{x}},\cdots,\vec{x}^{(n)}),
\end{equation}
where $\vec{x}$ is the trajectory of vortex center. Given the
anisotropic Hamiltonian $H=H_{2D}+\fft{\rm
g^2}{16\lambda^0}\int{d^2\xi}\delta (S^3)^2$, the
Eq.(\ref{ll_aniso}) can be rewritten as
\begin{equation}
\fft{\partial\vec{S}}{\partial\xi^0}=-\vec{S}\times\fft{\partial
H}{\partial \vec{S}},
\end{equation}
which yields an $(n+1)$-th order differential equation for
$\vec{x}(\xi^0)$.  We direct the readers to the Ref.\cite{MB:1999}
for more discussion on dynamics of those excitations.

\subsection{Spin chain from partial gauge fixing}
In the previous discussion we have succeeded in establishing the
correspondence between the 2D Heisenberg model and a rotating
membrane, here we would like to explore another possible
correspondence associated with the membrane.  In the case of
partial gauge fixing, one may instead obtain the XXX$_{1/2}$ spin
chain from the same limit. To illustrate this, let us consider the
following {\it partial} gauge fixing,
\begin{equation}
X^0=\kappa \xi^0,\qquad X^{10}=c\xi^1,
\end{equation}
and the determinant decomposes into
\begin{equation}
|G_{ij}|=|\tilde{G}_{ij}|+c^2g_{[10][10]}\tilde{G}_{22},
\end{equation}
where the $\tilde{G}_{ij}$ is now the pull back metric of subspace
$\{X^1,\cdots,X^9\}$. Then the action becomes
\begin{eqnarray}
&&S_{m_2}=\int{d^3\xi}\{\fft{1}{4\lambda^0}[\tilde{G}_{00}-(2\lambda^0T_2)^2|\tilde{G}_{ij}|-{\rm
g}^2\tilde{G}_{22}]+ T_2
b_{\mu\nu\lambda}\partial_0X^{\mu}\partial_1X^{\nu}\partial_2X^{\lambda}\},\nonumber\\
&&{\rm g} \equiv 2\lambda^0 c T_2\sqrt{g_{[10][10]}}.
\end{eqnarray}
In the tensionless limit, i.e. $T_2\to 0$ but $\rm g$ kept finite,
one maps the membrane to the target space $S^3$ and takes the fast
spinning limit as before.  It is not difficult to identify the
resulting action with the Heisenberg XXX$_{1/2}$ spin chain and
the Hamilton is given by
\begin{equation}
H_{1D}=\fft{\rm g^2}{16\lambda^0}\int{d\xi^2}
\partial_2\vec{S}\cdot \partial_2\vec{S}.
\end{equation}
Here we see that the membrane loses its dependance on $\xi^1$ and
behaves just like a string extending along $\xi^2$.  From the
viewpoint of {\it double} dimensional reduction where M-theory
reduces to type IIA string and membranes to strings, we might
think that the same spin chain system can also be obtained from
rotating D$1$ string inside $S^3$, in comparison with the case of
fundamental string\cite{Kruczenski:2003gt}. However, this is
puzzling since there is no D$1$ brane in the IIA string.  A better
interpretation would be that a compactified D$2$ brane behaves
like a one-dimensional string.  We are also tempted to make a
connection with the Matrix model interpretation of
M-theory\cite{Banks:1996vh}, where the dynamics is translated into
$(1+0)$ Matrix quantum mechanics and degrees of freedom are those
of D$0$ particles.  Here we conjecture that those D$0$ particles
form a chain and each of them possesses a spin on $S^2$.

\section{Heisenberg model/$p$-brane correspondence}
\subsection{$p$-dimensional $SU(2)$ spin model from rotating
$p$-brane} In principle, one may start with a rotating $p$-brane
inside a $S^3$ out of $D$-dimensional spacetime.  With gauge
fixing as follows\footnote{We always make the assumption that
there is enough space for a $S^3$ after gauge fixing.}:
\begin{equation}\label{p-gauge}
X^0=\kappa \xi^0,\qquad X^{D-p}=c_p\xi^p,\cdots, X^{D-1}=c_1\xi^1.
\end{equation}
It is not difficult to show that in the tensionless limit and fast
brane limit as before, we reproduce the $p$-dimensional $SU(2)$
Heisenberg model with the action,
\begin{eqnarray}\label{p-Heisenberg}
&&S_{p}=\fft{1}{16\lambda^0}\int{d^{p+1}\xi} \{ 2\partial_t\alpha+
2\cos{\beta}\partial_t\gamma -\sum_{i=1}^p{\rm
g}_i{}^2[(\partial_i\beta)^2+\sin^2{\beta}(\partial_i\gamma)^2]
\},\nonumber\\
&&{\rm g}_i=2\lambda^0T_p \prod_{j=1}^p
(c_j\sqrt{g_{[D-j][D-j]}})/(c_i\sqrt{g_{[D-i][D-i]}}),
\end{eqnarray}
where the limit has been taken with $T_p\to 0, c_i\to \infty$ but
finite $\rm g_i$.  As for D$p$ branes, more care is needed for its
non-vanishing dilaton $\Phi$ and antisymmetric B-field $B_{mn}$
from the closed string sector\footnote{Technically speaking,
dilaton field and antisymmetric B-field are in the same NS-NS
sector as graviton, namely both left- and right-moving modes
satisfy Neveu-Schwarz periodic condition.} as well as a $U(1)$
gauge field $F_{mn}$ from the open string sector. A Polyakov-type
action is given by\cite{Bozhilov:2002sj}
\begin{eqnarray}
S_{Dp}=&&\int{d^{p+1}\xi}\fft{e^{-a\Phi}}{4\lambda^0}[G_{00}-2\lambda^iG_{0i}+(\lambda^i\lambda^j-\kappa^i\kappa^j)G_{ij}-(2\lambda^0T_{Dp})^2|G_{ij}|+2\kappa^i({\cal F}_{0i}-\lambda^j{\cal F}_{ji})]\nonumber\\
&&+T_{Dp}c_{\mu_0\cdots\mu_p}\partial_0X^{\mu_0}\cdots\partial_pX^{\mu_p},
\end{eqnarray}
where ${\cal F}_{mn}\equiv
B_{mn}+2\pi\alpha'F_{mn}=\partial_{[m}{\cal A}_{n]}$. Equations of
motion for those Lagrange multipliers $\lambda^i,\kappa^i$ give
the constraints,
\begin{eqnarray}\label{v-constraint}
&&G_{00}-2\lambda^jG_{0j}+(\lambda^i\lambda^j-\kappa^i\kappa^j)G_{ij}+(2\lambda^0T_{Dp})^2|G_{ij}|+2\kappa^i({\cal
F}_{0i}-\lambda^j{\cal F}_{ji})=0,\nonumber\\
&&G_{oj}-\lambda^iG_{ij}=\kappa^i{\cal F}_{ij},\nonumber\\
&&{\cal F}_{0j}-\lambda^i{\cal F}_{ij}=\kappa^iG_{ij}.
\end{eqnarray}
The choice of vanishing $\lambda^i,\kappa^i$ decouples the $\cal
F$ from the metric and gives us the same Heisenberg model as in
the Eq.(\ref{p-Heisenberg}).  Another nontrivial choice is to keep
nonzero $\kappa^i$.  In the following, for simplicity we will
discuss the case with only $\kappa^1\neq 0$ and ${\cal A}_0=0$.
The action then simplifies as
\begin{equation}
S_{Dp}=\int{d^{p+1}\xi}\fft{e^{-a\Phi}}{4\lambda^0}[G_{00}+\kappa^1\kappa^1G_{11}
-(2\lambda^0T_{Dp})^2|G_{ij}|]+T_{Dp}c_{\mu_0\cdots\mu_p}\partial_0X^{\mu_0}\cdots\partial_pX^{\mu_p}.
\end{equation}
With the gauge fixing (\ref{p-gauge}) and target space given in
the Eq.(\ref{target_metric}), the action becomes
\begin{equation}\label{action_deformed}
S_{Dp}=\int{d^{p+1}\xi}\fft{e^{-a\Phi}}{16\lambda^0} \{
2\kappa\partial_0\alpha+ 2\cos{\beta}\kappa\partial_0\gamma
+\kappa^1\partial_0{\cal A}_1-\sum_{i=2}^p{\rm
g}_i^2(\fft{\kappa^1}{2\kappa}{\cal F}_{1i})^2-\sum_{i=1}^p{\rm
g}_i{}^2[(\partial_i\beta)^2+\sin^2{\beta}(\partial_i\gamma)^2]
\},
\end{equation}
where those constraints in (\ref{v-constraint}) have been used to
replace $\alpha$ and $G_{11}$. We should take $\kappa^1\sim
\kappa$ in order for ${\cal F}$ to survive the fast limit as
$\kappa\to \infty$. After the Legendre transformation, one obtains
the Hamiltonian of Heisenberg model with an extra {\it magnetic}
field ${\cal F}_{1i}$, that is,
\begin{equation}
H_{Dp}=\fft{1}{16\lambda^0}{\rm
g}^2\int{d^p\xi}e^{-a\Phi}[(\partial\vec{S})^2+\fft12 {\cal
F}_{1i}{\cal F}^{1i}].
\end{equation}
Here introduction of the nontrivial dilaton field and two-form
flux complicates the original Landau-Lifshitz equation.  We simply
comment that the effective coupling now varies with location and
an external magnetic field goes through the $p$-volume. Their
functions are to be determined by the modified equations of motion
derived from the action (\ref{action_deformed}).

\subsection{Lower dimensional Heisenberg model from partial gauge fixing}
In the previous section, we have learnt that with partial gauge
fixing, a rotating membrane can also have its correspondence in
the spin chain.  The same technique can easily apply to
$p$-branes, where a lower dimensional Heisenberg model is
obtained.  Here we simply mention the scheme without detail.
First we partially fix the gauge,
\begin{equation}
X^0=\kappa \xi^0,\qquad X^{D-q}=c_q\xi^q,\cdots, X^{D-1}=c_1\xi^1,
\end{equation}
where $0 < q < p$.  Then the determinant decomposes into
\begin{equation}
|G_{ij}|=|\tilde{G}_{ij}|+\prod_{j=1}^q
(c_j^2g_{[D-j][D-j]})/(c_i^2g_{[D-i][D-i]})|\tilde{G}_{jj}|+\cdots,
\end{equation}
where the $\tilde{G}_{ij}$ is the pull back metric of subspace
$\{X^1,\cdots,X^{D-q-1}\}$.  The rest terms on the right hand side
is irrelevant after both tensionless and fast limit are taken.
Then it is straightforward to show that $q$-dimensional Heisenberg
model can be derived.

\subsection{Anisotropy and external magnetic field}

As shown in the Ref.\cite{Wen:2006fw} and summarized in the
Appendix, similarly one can also deform the target $S^3$ for the
$p$-brane.  This corresponds to the $p$-dimensional anisotropic
$SU(2)$ Heisenberg model.  In addition to anisotropy, one may
wonder if a Zeeman term can be added for interaction with the
enternal field.  We have just demonstrated that a local coupling
and magnetic flux occurs typically for the D$p$-branes.  However,
this magnetic flux is somehow expected from the full spectrum of
string theory and takes arbitrary form unless specified by
equations of motion. In fact, there is an alternative way to
switch on external magnetic field, say along $\it z$-direction,
for both string and general $p$-branes. This is done $\it
geometrically$ by mixing two angles of $S^3$ unevenly. Start with
the usual parametrization,
\begin{equation}\label{s3}
ds^2=d\theta^2+\sin^2{\theta}d\psi_1^2+\cos^2{\theta}d\psi_2^2,
\end{equation}
and then perturb the mixing matrix with a deformation parameter
$b$, such that
\begin{eqnarray}
\left(\begin{array}{c}\theta\\ \psi_1\\ \psi_2 \end{array}\right)
=\left(
\begin{array}{ccc}
1/2&0&0\\
0&1/2(1+b)&-1/2\\
0&1/2(1-b)&1/2\end{array}\right)\left(
\begin{array}{c}\beta\\ \alpha\\ \gamma\end{array}\right).
\end{eqnarray}
The metric after transformation becomes
\begin{equation}
ds^2=\fft14[d\beta^2+d\alpha^2+d\gamma^2+2\cos{\beta}d\alpha
d\gamma-2b\cos{\beta}d\alpha^2-2bd\alpha d\gamma+b^2d\alpha^2].
\end{equation}
After taking the fast limit, in the same spirit of the small
deformation limit, we then send $b\to 0$ but only keep $\kappa^2b$
finite.  This results in an additional term $-2b\cos{\beta}$ to
the action (\ref{p-Heisenberg}), here $\kappa$ has been absorbed
into redefinition of $\kappa\dot{X}\to \dot{X}$ and $\kappa^2 b\to
b$. It can be further put into the form $-\vec{B}\cdot\vec{S}$ and
precisely interpreted as the interaction with external magnetic
field $\vec{B}=2b\hat{z}$. This uneven mixing of angles was
already observed in \cite{Harmark:2006ie}, there appears as mixing
of angular velocities.  In fact, it can be shown that our double
scaling limit of small deformation is equivalently to their
decoupling limit.

\section{Conclusion}

We have realized in the fast membrane limit the 2D Heisenberg
model, which is integrable at least for the ansatz of travelling
wave.  In the Ref.\cite{Bozhilov:2007mb}, the author found several
types of membrane embedding into the $AdS_4\times S^7$ background,
which are related to the Neumann and Neumann-Rosochatius
integrable systems.  Our result supports his conjecture that there
might have various kinds of integrable system dual to membranes in
the M-theory.

In this paper, we also proposed the correspondence between
$p$-branes and $p$-dimensional Heisenberg spin model.  For
D$p$-branes, in particular, the nontrivial dialton field generates
a position-varying coupling and the two-form flux couples to the
system magnetically.  Later we also provide a geometric
realization of external magnetic field in the $SU(2)$ Heisenberg
model in arbitrary dimension.

Several directions may deserve further investigation. Here we only
mention some of them:  In the string/spin chain correspondence,
one is able to identify the anomalous dimension of single trace
operator $\Tr(ZZ...Z)$, calculated from the dual super Yang-Mills
theory, with the Hamiltonian of spin chain. The two or higher
dimensional spin model may not have such a correspondence due to
lacking the knowledge of {\it hyper}surface operators.  However,
the spin chain model obtained from partial gauge fixing can still
have its correspondence on the dual field theory side. In the
present case with membranes, calculation of single trace operators
in a three-dimensional ${\cal N}=8$ super Yang-Mills (CFT$_3$ as
its IR fixed point) may support this correspondence.

In the view of $2$D lattice, it is tempted to also switch on
interaction among diagonal sites, which counts as next-nearest
neighbors as previously mentioned.  It could be interesting to
investigate the way it appears in the worldvolume action, in
comparison to the $\alpha'$ correction in the string worldsheet.

Finally, in the Ref.\cite{Bjornsson:2004yp}, we have learnt that
under a partial gauge fixing, a membrane can be seen as a
perturbation around string like configuration, where the membrane
tension acts like the coupling constant.  It is tempting to relax
our tensionless limit and study the deviation from the Heisenberg
model perturbatively.

\section{acknowledgements}  I am grateful to invitation of NCTS/NTHU to present this work at its final stage.
I would like to thank K.~Furuuchi, P.~M.~Ho,
S.~Teraguchi, D.~Tomino, Q.~S.~Yan, and S.~Zeze for useful
discussion and comments. The author is supported in part by the
Taiwan's National Science Council under Grant No.
NSC95-2811-M-002-013.

%%%%%%%%%%%%%%%%%%%%%%%%%%%%%%%%%%%%%%%%%%%%%%%%%%%%%%%%%%%%%%%%%%%%%%%%%%%%
\appendix
\section{Anisotropic Heisenberg model from deformed $S^3$}

The deformed metric is given by
\begin{equation}\label{metric_deform}
ds^2=\fft{1}{4}[-dt^2+d\beta^2+\sin^2{\beta}d\alpha^2+(1-\delta)(d\gamma+\cos{\beta}d\alpha)^2]
\end{equation}
We request $0\le \delta \le 1$ to avoid non-unitary gauge field
($\delta<0$) and closed time-like geodesics ($\delta>1$).  For
trivial $\delta=0$, one recovers the unit round $S^3$ as Hopf
fibration. For maximal $\delta=1$, the $S^1$ fiber degenerates and
we are left with $S^2$.

After sending $\alpha\to \alpha+t$ and taking fast limit as the
undeformed one, then the relevant part becomes
\begin{eqnarray}\label{action_2D}
S_{m_2}'=&&\fft{1}{16\lambda^0}\int{d^3\xi} \{
-\delta\cos^2{\beta}\kappa^2+ 2\kappa\dot{\alpha}+
2\kappa\cos{\beta}\dot{\gamma} -\sum_{i=1}^2{\rm
g}_i{}^2[(\partial_i\beta)^2+\Delta_{\delta}(\partial_i\gamma)^2]
\},\nonumber\\
&&\Delta_{\delta}\equiv\fft{(1-\delta)\sin^2{\beta}}{1-\delta\cos^2{\beta}}.
\end{eqnarray}
In order to keep only the anisotropic term, we have to take the
{\it small deformation limit} by sending $\delta\to 0$ but keep
$\kappa^2\delta$ finite.  In this limit, the $\Delta_\delta\to
\sin^2{\beta}$ as desired.  After both fast and small deformation
limits are taken, it is convenient to rescale $\kappa^2 \delta \to
\delta$ and $\kappa\dot{X}\to\dot{X}$. The equations of motion for
$\beta$ and $\gamma$ are then given by
\begin{eqnarray}\label{eom}
&&{\rm g}^2\beta'' -
\sin{\beta}\dot{\gamma}-{\rm g}^2\sin{\beta}\cos{\beta}(\gamma')^2+\delta\sin{\beta}\cos{\beta}=0\nonumber\\
&&\sin{\beta}\dot{\beta}+{\rm g}^2(\sin^2{\beta}\gamma')'=0,
\end{eqnarray}
which gives rise to the general Landau-Lifshitz equation,
\begin{eqnarray}
&&\partial_t \vec{S} = {\rm g}^2\vec{S} \times \partial^2\vec{S} +
\vec{S}\times
{\cal J}\vec{S},\nonumber\\
&&{\cal J}_{11}={\cal J}_{22}=1,\qquad {\cal J}_{33}=1-\delta,
\end{eqnarray}
with the same spin vector
$\vec{S}=(\sin{\beta}\cos{\gamma},\sin{\beta}\sin{\gamma},\cos{\beta})$.

This general Landau-Lifshitz equation can be seen as continuous
limit of the inhomogeneous Heisenberg spin chain model.

\end{document}